# Decreasing Digital Distraction in College Students: Associated Online Learning Strategies Identified by Unsupervised Data Mining Approaches

Hui Shi[1], Ran Bi[2], Xi Lin[3], and Yan Dai[4]

[1]Faculty of Artificial Intelligence in Education, Central China Normal University, Wuhan, China
[2]Department of Advanced Analytics Research and Development, SAS Institute Inc, Cary, NC, USA
[3]Department of Interdisciplinary Professions, East Carolina University, Greenville, NC, USA
[4]Department of Psychology & Sociology, Tuskegee University, Tuskegee, AL, USA

## Abstract

The proliferation of digital tools in education offers numerous benefits but also introduces significant challenges, notably digital distractions that hinder academic performance, especially in online learning contexts. This study employed unsupervised data mining techniques, specifically association rule mining and clustering analysis, to identify effective learning strategies associated with lower levels of digital distractions among college students. Data from 530 participants revealed that self-regulated learning strategies (i.e., goal setting, environment structuring, and time management) co-occurred most consistently with lower digital distractions. Additionally, learner-instructor and learner-content engagement strategies, as well as technical competencies, also tended to appear in the same profiles as lower distraction. Interestingly, reliance on peer help-seeking and learner-learner engagement strategies appeared less often in those lower distraction profiles. These findings offer actionable implications for educators to design targeted interventions that foster focused and productive online learning environments.

*Keywords*: digital distractions; self-regulated learning; engagement; online learning readiness; distance education and online learning; association rule mining; machine learning

## Introduction

In contemporary education, the reliance on digital technologies is ubiquitous, particularly intensified by the global shift toward online learning precipitated by the COVID-19 pandemic (Aristovnik et al., 2023; Getenet et al., 2024; Rivers et al., 2022; Wu & Xie, 2018). Laptops, tablets, and smartphones have become indispensable tools for students, facilitating communication, collaboration, and access to learning resources (Abed & Shackelford, 2024; Campbell et al., 2024). Learning management systems such as Canvas and Blackboard streamline course delivery and performance tracking, along with online platforms such as MOOCs making education more inclusive by providing access regardless of geographical or economic barriers (Bradley, 2021; Wang et al., 2022).

However, this digital transformation is accompanied by a growing concern: digital distraction, defined as off-task behaviors triggered by the pervasive presence of digital devices (Agrawal et al., 2017), often involving the shift from academic tasks to unrelated online activities such as social media or gaming (Kraushaar & Novak, 2010). Research has demonstrated its pervasive nature in educational contexts. Rosen (2017) reported that college students checked their phones approximately 50 times daily and had 4.5 hours of screen time, often disrupting their ability to focus on academic tasks. The issue is especially acute in online learning environments where constant access to the Internet and limited external monitoring facilitate multitasking (e.g., texting, social media browsing, and internet surfing) and off-task device usage (Aivaz & Teodorescu, 2022; Nabung, 2024).

Previous research has predominantly examined demographic and contextual factors contributing to digital distraction (e.g., age, gender, instructional design) (Chen et al., 2014; Baturay & Toker, 2015; Taneja et al., 2015). While informative, such studies focus mainly on identifying correlates rather than proposing empirically grounded solutions. As a result, there remains a significant research gap in identifying and validating specific learning strategies and competencies that effectively mitigate digital distraction (Deng et al., 2024). Second, existing research primarily addresses digital distractions in face-to-face contexts, leaving a gap concerning online learning environments (Göl et al., 2023). Critically, students may not rely on a single learning strategy but typically employ multiple strategies simultaneously throughout their learning processes. However, few studies have explored how these strategies co-occur or interact, particularly among students who demonstrate lower levels of digital distraction, leaving the interconnections among strategies under-investigated.

To address these gaps, this study employed advanced unsupervised data mining methods, specifically, association rule mining (ARM) and clustering analysis, to uncover patterns and interrelationships among online learning strategies that are associated with digital distraction. Unlike traditional inferential statistics, such as regression, which require predefined hypotheses, assumptions of normality, and sufficient sample sizes to ensure statistical power and validity, as well as their focus on variable-specific effects, unsupervised machine learning techniques offer a more flexible and exploratory alternative. ARM is particularly effective in uncovering frequent and co-occurring patterns among variables without imposing rigid assumptions about data distribution or linear relationships (Hahsler et al., 2005; Haque et al., 2024). This enables the detection of multi-dimensional interactions that might otherwise go unnoticed using traditional approaches. Besides, clustering analysis complements ARM by grouping the discovered patterns into meaningful themes, offering practical insights into different profiles of learners. Thus, this combined approach is especially well-suited for investigating how multiple learning strategies might co-occur among students and how those co-occurring strategies are statistically associated with different levels of digital distraction.

Therefore, the objectives of this study are twofold: 1) to identify online learning strategies that are significantly associated with lower levels of digital distraction, and 2) to explore the underlying structure and interrelationship of these strategies using clustering of association rules. Based on existing literature, we hypothesized that college students' digital distraction in online learning would be significantly associated with their self-regulated learning strategies, engagement strategies, and online learning readiness. The theoretical rationale and empirical evidence supporting this hypothesis are elaborated in the following literature review sections. To clearly articulate the objectives and contributions of this research, the following questions are posed:

RQ1: What online learning strategies are significantly associated with lower levels of digital distraction among college students in online courses?

RQ2: How do these strategies interrelate to collectively influence digital distraction?

# Literature Review

## Digital Distraction in Higher Education

The widespread adoption of digital technologies in higher education has provided numerous benefits, enhancing student engagement, attentiveness, and study efficiency (Samson, 2010). However, the unclear boundary between productive and distractive media use often contributes to attention problems (Kraushaar & Novak, 2010). Sun and Chao's (2024) research, which was conducted among 887 students, found that these students frequently multitask between learning and socializing via social media for entertaining purposes. They thereby concluded that media multitasking often leads to attention problems, negatively impacting academic performance. In short, media multitasking can intensify digital distractions (Göl et al., 2023).

Researchers have identified personal and course-related factors influencing digital distractions. Males typically checked digital devices more frequently than females during classes (Baturay & Toker, 2015), yet digital distraction, specifically mobile phone distraction, harms female students significantly more than male students (Chen et al., 2025). Younger students were generally more susceptible to such distractions (Martin et al., 2025). Additionally, instructional design influences digital distraction (Flanigan & Babchuk, 2022); poorly designed online content can increase non-academic digital use or off-task activities, described as ‘cyber-slacking’ (Taneja et al., 2015, p. 141) or cyber-loafing (Gerow et al., 2020; Dang et al., 2024).

However, simply identifying these factors is insufficient; effective strategies to manage distractions should also be investigated. Educators need to equip students with targeted learning strategies to regulate attention effectively in digital environments (Flanigan & Babchuk, 2022).

Firstly, engaging interactions between learners, instructors, and content help minimize distractions. Peer-to-peer interactions promote collaborative learning and accountability, thus limiting distractions (Flanigan & Babchuk, 2022). Instructor-student interactions, through interactive and participatory teaching methods, reduce student apathy and enhance attention (Crawford et al., 2020). Similarly, active learning approaches and relatable course content foster student interest and focus (Flanigan & Babchuk, 2022; Martin & Bolliger, 2018).

Secondly, researchers recommended teaching students self-regulated learning strategies, such as setting clear goals and prioritizing tasks, to manage and maintain attention effectively (Taghavi-Nejad et al., 2024; Yu-Lin et al., 2022). Additionally, fostering metacognitive skills enables students to consciously monitor and regulate their digital device usage (Rosen, 2017).

Further, effectively using metacognitive strategies has the potential to positively influence students' learning activities (Lin et al., 2024).

Finally, the recent shift toward online education underscores the need for technical, social, and communication competencies to reduce digital distractions (Göl et al., 2023; Masry-Herzallah et al., 2024). These skills are closely related to self-regulated learning, empowering students to utilize digital platforms productively (Dai et al., 2023). Additionally, developing ICT self-efficacy and resilience is essential for students to adapt successfully in technology-mediated learning environments (Hatlevik & Bjarnø, 2021).

Thus, reducing digital distraction in higher education necessitates a deeper understanding of the learning strategies that foster focused engagement. In the following sections, the literature review focuses on three theoretical and empirical domains: self-regulated learning (SRL), engagement strategies, and online learning readiness (OLR). It further identifies underexplored interconnections among these constructs in relation to digital distraction.

**Self-regulated Learning Strategies and Digital Distraction**

SRL refers to learners' proactive processes in managing cognitive, motivational, and behavioral aspects to achieve learning goals, encompassing phases of forethought, performance, and self-reflection (Pintrich, 2004; Zimmerman, 2000). SRL strategies such as goal setting, environment structuring, task and time management, help-seeking, and self-evaluation, are essential in online contexts, given the higher requirement for learner autonomy and limited direct instructor supervision (Barnard et al., 2009; Shi et al., 2024).

These strategies not only foster academic success but are also instrumental in reducing distractions. For example, goal setting and planning help students prioritize tasks and allocate time effectively (Wang et al., 2022); environment structuring, such as silencing notifications and designating distraction-free zones, creates physical barriers to distraction (Deng, 2020); and self-evaluation allows students to reflect on off-task behaviors and improve future performance (Wang et al., 2022). While these associations are theoretically sound, empirical investigations on the specific SRL strategies most effective in reducing digital distraction are scarce (Deng et al., 2024), especially within online learning environments (Göl et al., 2023).

## Engagement Strategies and Digital Distraction

In online learning, engagement is often operationalized through learner-instructor, learner-learner, and learner-content interactions (Moore, 1989; Martin & Bolliger, 2018). Well-designed online course activities, such as using memes (Lin & Sun, 2023), video-timeline-anchored discussions (Lin et al., 2024), and role play with artificial intelligence (Lin et al., 2024), have been shown to mitigate disengagement and apathy (Lin & Sun, 2024). Such activities could limit distraction (Lund et al., 2012) and enhance cognitive engagement and reduce the tendency toward off-task behaviors (Khan et al., 2017; Muir et al., 2022). Engagement strategies are empirically supported in maintaining student motivation in online courses (Martin & Bolliger, 2018), but whether and how they influence distraction remains ambiguous, suggesting the need for further analysis.

## Online Learning Readiness and Digital Distraction

OLR encompasses the technical, social, and communication competencies needed for effective online learning (Yu, 2018). Students who frequently use personal technology may face increased risks of digital distractions during online or blended learning (Kumar et al., 2024). On the opposite, students with higher competencies to use technologies effectively for academic purposes might be more likely to stay focused and on-task. Nevertheless, very limited work has examined whether and how students' OLR influences digital distraction. In addition, previous literature has highlighted an interrelationship between OLR and SRL. Sahdan et al. (2017) and Lin and Dai (2022) found that students better prepared for online learning exhibit significantly stronger SRL behaviors, including more effective time management and help-seeking. However, literature insufficiently addresses how these interrelationships influence digital distractions. Thus, there remains an empirical gap concerning whether and how OLR and SRL competencies jointly affect digital distractions in online contexts.

# Methods

## Participants

We adopted convenience sampling due to practical constraints in accessing a large, diverse pool of online learners and to efficiently recruit participants with direct experience in synchronous online courses, which aligned with the focus of our study. Participants in this study were drawn from a public university in southeastern America, where full-time college students were recruited to complete an online survey. Only participants who had attended at least one synchronous online course at a university were eligible to participate in the survey. They were asked to reflect on one specific course or multiple synchronous online courses they had studied before completing the survey. The participants provided consent before completing the survey. The author's university Institutional Review Board approved all study protocols.

Initially, A total of 764 students completed the online survey. Following data cleaning procedures, 96 incomplete responses were removed, along with 6 unengaged responses that showed uniform answers across all questions. 103 responses were excluded due to incorrect completion of required items (e.g., Please choose "somewhat agree" on this question), and 29 responses with a completion time of 5 minutes or less were also deleted. After these procedures, 530 valid responses remained for analysis. Table 1 summarizes demographic characteristics of students. About 91% of undergraduate students, averaging 22 years old (males = 25%) from diverse majors, participated in this study. Although the sample included a higher proportion of female participants, an independent samples t-test indicated that no significant difference in digital distraction between male and female students (t = 0.86, $p$ = .39), suggesting that gender imbalance in the sample is unlikely to bias the findings. This conclusion is further supported by the association rule mining results, which did not identify gender as a meaningful antecedent attribute when characterizing lower levels of distraction. Details can be found in the following sections.

**Table 1**

*Participant Characteristics*

| | Full sample (n = 530) | Digital Distraction | | $\chi^2$(df) / t-test | $p$ |
|---|---|---|---|---|---|
| | | Low (n = 216) | High (n = 314) | | |
| Age in years, Mean (SD) | 21.63 (5.39) | 22.06 (6.32) | 21.33 (4.64) | 1.52 (528) | .130 |
| Gender, N (%) | | | | 0.01 (1) | .935 |
| Male | 131 (24.7%) | 54 (25.0%) | 77 (24.5%) | | |
| Female | 392 (74.0%) | 160 (74.1%) | 232 (73.9%) | | |
| Race, N (%) | | | | 11.66 (7) | .112 |
| White | 365 (68.9%) | 158 (73.1%) | 207 (65.9%) | | |
| Black | 32 (6.0%) | 13 (0.06%) | 19 (6.1%) | | |
| Hispanic | 69 (13.0%) | 26 (12.0%) | 43 (13.7%) | | |
| American Indian | 2 (0.4%) | 0 | 2 (0.6%) | | |
| Asian | 25 (4.7%) | 8 (3.7%) | 17 (5.4%) | | |
| Native Hawaiian | 1 (0.2%) | 0 | 1 (0.3%) | | |
| Two or more races | 29 (5.5%) | 6 (2.8%) | 23 (7.3%) | | |
| Grade, N (%) | | | | 1.15 (5) | .886 |
| Freshman | 26 (4.9%) | 10 (4.6%) | 16 (5.1%) | | |
| Sophomore | 106 (20.0%) | 45 (20.8%) | 61 (19.4%) | | |
| Junior | 164 (30.9%) | 62 (28.7%) | 102 (32.5%) | | |
| Senior | 188 (35.5%) | 79 (36.6%) | 109 (34.7%) | | |
| Graduate | 45 (8.5%) | 20 (9.3%) | 25 (8.0%) | | |
| Non-degree seeking | 1 (0.2%) | 0 | 1 (0.3%) | | |
| Major, N (%) | | | | 1.18 (6) | .978 |
| Arts & Humanities | 55 (10.4%) | 22 | 33 | | |
| Business | 90 (17.0%) | 39 (18.1%) | 51 (16.2%) | | |
| Education | 68 (12.8%) | 26 (12.0%) | 42 (13.4%) | | |

| | | | |
|---|---|---|---|
| Health Sciences | 109 (20.6%) | 47 (21.8%) | 62 (19.7%) |
| Psychology | 78 (14.7%) | 29 (13.4%) | 49 (15.6%) |
| Science & Engineering | 46 (8.7%) | 18 (8.3%) | 28 (8.9%) |
| Social Sciences | 84 (15.8%) | 35 (16.2%) | 49 (15.6%) |

**Measures**

The measures collected in this study included data related to student demographics, perceptions of engagement strategies in attending online courses, OLR, online SRL strategies, and student reports of digital distraction activities in online classes. All scales were used in their original validated form without major modifications. Minor wording adjustments were made for contextual clarity, but did not alter the meaning of the items.

A confirmatory factor analysis (CFA) was conducted to verify whether the factor structures of the scales used in this study aligned with the collected data. CFA assesses whether the models proposed by the data align with theoretical assumptions and whether the hypothesized relationships in the theoretical population are present in the empirical dataset (Anderson & Gerbing, 1988). The structures of all the survey scales were determined using the unweighted least squares (ULS) method. Specifically, we used the estimator of ULSMV, where both the mean and variance of chi-square are adjusted, because ULSMV performs better with ordinal data and smaller sample sizes than other estimation methods (Forero et al., 2009; Xia & Yang, 2019). Model fit indexes were analyzed using the chi-square to degrees of freedom ratio ($\chi^2$/df), root mean square error of approximation (RMSEA), comparative fit index (CFI), Tucker-Lewis index (TLI), and standardized root mean squared residual (SRMR). The values of $\chi^2/df < 4$, RMSEA < .08, CFI and TLI > .90, and SRMR < .08 are considered an acceptable fit (Hu & Bentler, 1999). Detailed information regarding questionnaires used in this research is provided below.

*Engagement Strategies in the Online Learning Environment Questionnaire (ESOQ)*

The ESOQ was constructed and validated by Martin and Bolliger (2018) to investigate college students' perceptions of the importance of engagement strategies when attending online courses. The scale is a 5-point Likert scale anchored with 1 = very unimportant for me and 5 = very important for me, measuring three dimensions of engagement strategies based on Moore's (1993) three types of interaction framework: Learner-to-Instructor Engagement (LI), Learner-to-Learner Engagement (LL), and Learner-to-Content Engagement (LC). All calculated Cronbach's alpha coefficients measuring internal consistency for the ESOQ factor within the survey were above 0.8. CFA results showed that the three-dimensional model of the instrument generally has a good fit [$\chi^2$/df = 3.24, RMSEA = .065, CFI = .915, TLI = .905, SRMR = .058].

*Online Learning Readiness Questionnaire (OLRQ)*

OLRQ was developed by Yu (2018), including items measuring four factors: (a) Technical Competency (TC), (b) Social Competency with Instructor (SCI), (c) Social Competency with Classmates (SCC), and (d) Communication Competency (CC). This survey consists of 20 items and was measured on a 5-point Likert scale ranging from 1 = strongly disagree to 5 = strongly agree. The TC factor investigates students' experience with computer technologies and their confidence in the process of integration. SCI surveys self-efficacy for interacting with instructors in an online course. SCC includes items to survey students' self-efficacy for interacting with classmates in an online course. CC includes items measuring communication competencies in online learning. All calculated Cronbach's alpha coefficients calculated in this questionnaire were higher than 0.8. Additionally, CFA results indicated that the OLRQ fits the four-dimension measurement model [$\chi^2$ /df = 2.53, RMSEA = .054, CFI = .955, TLI = .948, SRMR = .044].

*Online Self-Regulated Learning Questionnaire (SRLQ)*

SRLQ is a 19-item 5-point Likert scale, constructed and validated by Barnard et al. (2009) evaluating SRL strategies with college students in online or blended learning environments. This research used the following subscales of online self-regulated strategies: goal setting, environment structuring, task and time management strategies, help-seeking, and self-evaluations. The current research found that the reliability coefficients ranged between .637 and .858 for sub-dimensions of online SRL questionnaires. CFA results indicated that the questionnaire fits the five-dimension measurement model [$\chi^2$ /df = 2.27, RMSEA = .049, CFI = .931, TLI = .917, SRMR = .064].

*Digital Distraction Questionnaire (DDQ)*

DDQ is a 4-item instrument measured on a 5-point Likert scale ranging between 1 = never and 5 = very frequently (Chen et al., 2014). In the present research, the reliability of this scale is .814. CFA results revealed that fit indices for the model were acceptable [$\chi^2$/df = 3.96, RMSEA = .075, CFI = .992, TLI = .975, SRMR = .020].

**Data Analysis**

To answer the research questions, our data analysis was conducted in two stages, each employing different data mining techniques. The analytical process is depicted in Figure 1. There were two major analytical steps: 1) the first analytical step concentrated on using association rule mining to identify associated and significant learning strategies with digital distraction decrease; and 2) the second analytical step focused on discovered rules as a unit of analysis, using clustering to uncover broad patterns in rules with multiple attributes. All analyses were conducted using R version 4.2.3 with *arules* (Hahsler et al., 2005; Zhao, 2012) and *arulesViz* packages (Hahsler & Chelluboina, 2011). Each analytical step is illustrated in detail as follows (see Figure 1).

**Figure 1**

*Data Analytic Process.*

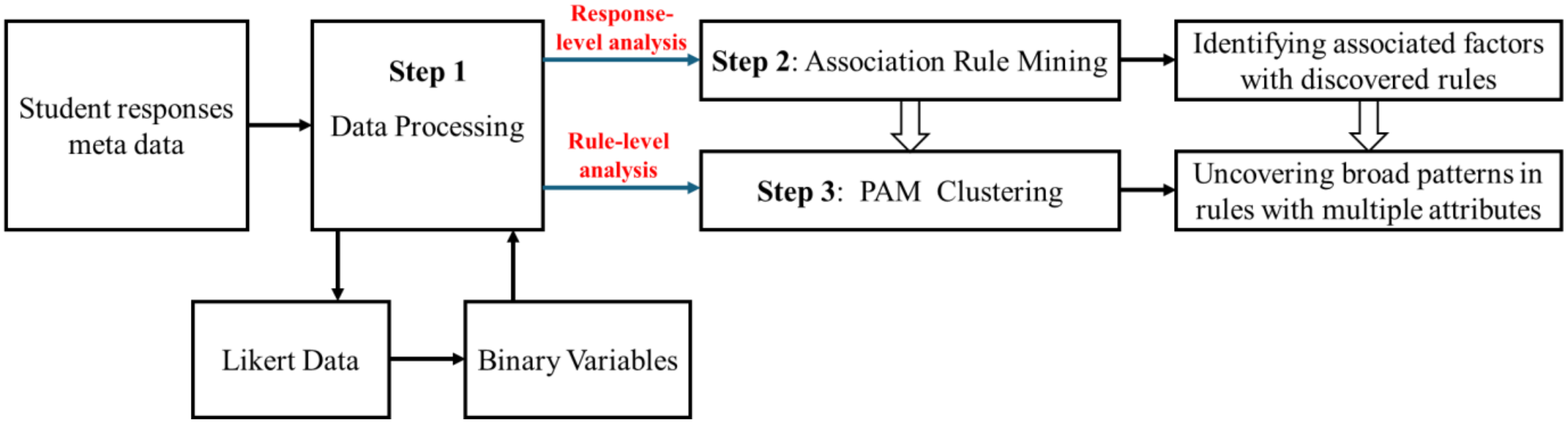

*Variables Dichotomization*

To perform ARM, it is necessary to convert all data into dichotomous variables, with each variable recorded as either present or absent for each participant. For instance, the variable "gender" was split into two variables: 1) female, where a female participant is coded as True and a male participant as False; and 2) male, where a male participant is coded as True and a female participant as False. A questionnaire item using a 5-point Likert scale would be transformed into five separate variables, each representing whether a participant endorsed a specific rating or not. Likert scales were further decomposed into two levels. For instance, ESOQ subscale questions used a 5-point Likert scale from Very Unimportant to Very Important; therefore, the responses were dichotomized into variables reflecting the Likert options and variables to capture high and low importance: greater than Neutral and less than Neutral. Regarding our outcome variable, we dichotomized digital distraction into high and low levels based on the participants' mean values. Thus, the dichotomization resulted in a total of 447 binary variables.

*Association Rules Mining*

ARM identifies frequent *itemsets*—combinations of dichotomous variables that co-occur within individual observations (i.e., participants)—and generates association *rules* in an "if-then" format (Hahsler et al., 2005). These rules, consisting of *antecedents* (the "if" part) and *consequents* (the "then" part), reveal how the co-occurrence of certain items or events predicts the outcome variables of interest. In this study, frequent itemsets were identified using the *Apriori* algorithm (Agrawal & Srikant, 1994) and were deemed *frequent* if they met or exceeded a specified threshold, known as *support*. The derived rules indicate that if itemset X occurs, then Y is likely to occur. Here, X represents the antecedent learning strategy attributes on the left-hand side of the rule, while Y, the consequent, is on the right-hand side. In this research, Y represented either a high or low level of digital distraction.

Instead of the interest in frequencies between antecedent and consequent attributes, we are more interested in the strength of associations, as it reveals meaningful and reliable patterns. To evaluate the strength of the association rules, four metrics were calculated: support, confidence, lift, and phi.

Confidence measures the reliability of a rule by representing the conditional probability that the consequent will occur given the presence of the antecedent (Agrawal et al., 1993). For instance, if X and Y co-occur in 10% of participants, but X alone appears in 70%, the confidence is low because Y is present in only a small subset of those with X. Lift evaluates the strength of a rule by comparing the observed co-occurrence of the antecedent and consequent to what would be expected if they were independent (Agrawal et al., 1993). A lift value greater than 1 indicates a positive association between the antecedent and the consequent, while a value less than 1 suggests a negative association. The phi coefficient, or Pearson's correlation coefficient, measures the correlation between the antecedent and consequent (Yan et al., 2009), ranging from

-1 to 1. A value of 1 indicates a perfect positive association, -1 indicates a perfect negative association, and 0 signifies no association.

$$Supp(X \rightarrow Y) = \frac{n(XY)}{n} \quad (1)$$

$$Conf(X \rightarrow Y) = \frac{supp(X \cup Y)}{supp(X)} \quad (2)$$

$$Lift(X \rightarrow Y) = \frac{supp(X \rightarrow Y)}{Supp(X)Supp(Y)} \quad (3)$$

$$phi\ (X \rightarrow Y) = \frac{supp(X \cup Y) - supp(X)supp(Y)}{\sqrt{supp(X)supp(Y)}(1 - supp(X))(1 - supp(Y))} \quad (4)$$

*Extracting Significant Rules*

ARM was used to identify associated online learning strategies that related to digital distraction decrease reported by students. To do this, lift was set to less than 0.95, and phi was set to less than 0 to extract the significant antecedent attributes (i.e., learning strategies) that have a negative association with digital distraction. We first extracted single-antecedent rules for characterizing low levels of digital distraction and then identified rules with two antecedents for better characterization. We also determined minimum support (20%) and minimum confidence (50%) thresholds for rules to ensure that (1) the patterns were common enough in the entire sample of 530 college students; (2) data was allowed for statistical testing ($n \geq 30$). The chi-square test was employed to assess whether the identified antecedent attributes disproportionately appeared among students with high or low digital distraction levels.

*Clustering of Rules*

To identify patterns among the generated rules based on multiple attributes, clustering was utilized, specifically, partitioning around medoids (PAM) was applied, as it has been demonstrated to handle association rules more effectively (Maechler et al., 2014). Normalized distances between rules were calculated as 1 minus the conditional probability, which measures how often the rules co-occurred in the same participant compared to how often they occurred independently (Gupta et al., 1999). Since the clustering was based on rules rather than participants, the resulting clusters can be interpreted as sets of learning strategies linked to low levels of digital distraction across overlapping participant subgroups. This implies that a participant could be represented by one cluster or multiple clusters of related characteristics. Consequently, each cluster of rules can be interpreted as a hypothesized group of characteristics or indicators associated with digital distraction decrease. We used a combination of the Elbow Method and Silhouette scores to determine the optimal number of clusters (k) (Rousseeuw, 1987). We also evaluated internal validity and cluster stability to ensure the clusters were distinct and meaningful (Brock et al., 2008).

## Results

### Descriptive Statistics

The study involved 530 college students who participated in this online survey. Participants ranged in age, with an average of 21.63 years, predominantly female (74.0%), and mostly White (68.9%). The sample included freshmen through graduate students, with seniors being the largest subgroup (35.5%) (see Table 1). Participants were categorized into high digital distraction (N = 314, 59.2%) and low digital distraction (N = 216, 40.8%) groups by performing a median split (Iacobucci et al., 2015). No significant demographic differences were found between these groups ($p > .05$).

### Associated Strategies for Digital Distraction Decrease

#### *Step 1: Single Antecedent Attribute in ARM*

The initial step of ARM identified 19 significant rules with single attributes (see Table 2), revealing negative associations (lift values ranging from 0.89 to 0.95, phi < 0) between antecedent attributes and digital distractions. These results suggested that specific online learning strategies were associated with lower levels of distraction. Notably, four redundant rules were excluded, as they provided no additional information over existing ones (Kingir et al., 2020). For instance, the rule OLRQ 10 = Strongly Agree ("I am confident I can timely inform the instructor when unexpected situations arise") overlapped with the broader rule OLRQ 10 > Neutral and was subsequently removed.

Among significant rules, SRL strategies appeared prominently associated with lower distraction levels (9 rules, see GS, ES, TMTS rules in Table 2). Specifically, rules related to goal setting, task and time management, and environment structuring indicated significant associations. Students who reported lower digital distractions also tended to report confidence in setting clear goals to manage their study time and maintain consistent quality in their work across the week. Furthermore, students with fewer distractions tended to describe choosing study times with minimal interruptions and keeping high standards for their learning.

Furthermore, engagement strategies with instructors and learning content also emerged as negatively associated with digital distractions (see LI and LC rules in Table 2). For example, instructor-driven reflective opportunities and interactive synchronous sessions correlated with lower distraction levels. Similarly, the use of optional online resources to explore topics in greater depth also correlated with lower distraction levels. In addition, students' technical confidence using digital learning tools was positively associated with lower distraction levels.

Interestingly, some rules (see HS rules in Table 2) revealed that students experiencing fewer distractions reported less frequent peer-related help-seeking behaviors, such as sharing problems with classmates or seeking help from knowledgeable peers.

We found no significant rules that included student demographic characteristics (i.e., gender, age, race) and majors, which filtered out demographics as meaningful predictors because they did not co-occur with lower distraction patterns under our significant rule extraction thresholds (support > .20 and confidence > .50). Thus, demographics did not play a role in the behavioral profiles that characterize lower distraction.

In sum, these associative findings suggest that lower digital distraction levels tend to co-occur with SRL behaviors, meaningful engagement with instructors and content, and confidence in technical competencies. Peer interaction strategies, notably help-seeking, appeared inversely associated with mitigating digital distractions.

**Table 2**

*Individual Learning Strategies that Co-occur with Lower Digital Distraction*

| Subscale | Attribute | N | Supp. | Conf. | Lift | Phi | Odds Ratio | $p^a$ | Interpretation |
|---|---|---|---|---|---|---|---|---|---|
| LI | ESOQ 14 = Agree | 115 | 0.217 | 0.550 | 0.929 | -0.069 | 0.750 | .111 | Students who agreed that the instructor creates a clear course orientation for students were more likely to be in the low digital distraction group. |
| LI | ESOQ 18 = Agree | 115 | 0.217 | 0.558 | 0.942 | -0.055 | 0.794 | .201 | Students who agreed that the instructor provides students with an opportunity to reflect were more likely to be in the low digital distraction group. |
| LI | ESOQ 20 = Agree | 108 | 0.204 | 0.543 | 0.916 | -0.078 | 0.720 | .071 | Students who agreed that the instructor uses various features in synchronous sessions to interact with students were more likely to be low distraction. |
| LC | ESOQ 22 = Agree | 152 | 0.287 | 0.545 | 0.920 | -0.102 | 0.658 | .019 | Students who agreed that they use optional online resources to explore topics in more depth were more likely to be low distraction. |
| LC | ESOQ 28 = Agree | 124 | 0.234 | 0.530 | 0.894 | -0.113 | 0.629 | .009 | Students who agreed that they work on realistic scenarios to apply content were more likely to be low distraction. |
| TC | OLRQ 1 = Agree | 150 | 0.283 | 0.558 | 0.941 | -0.072 | 0.746 | .098 | Students who felt confident using computer technologies |
| | | | | | | | | | for specific tasks were more likely to be low distraction. |
| SCI | OLRQ 10 > Neutral | 223 | 0.421 | 0.560 | 0.946 | -0.114 | 0.574 | 009 | Students who felt confident that they can timely inform the instructor when unexpected situations arise were more likely to be low distraction. |
| SCI | OLRQ 11 > Neutral | 181 | 0.342 | 0.562 | 0.949 | -0.077 | 0.724 | .077 | Students who felt confident that they can express their opinions to the instructor respectfully were more likely to be low distraction. |
| SCC | OLRQ 12 < Neutral | 151 | 0.285 | 0.559 | 0.944 | -0.069 | 0.755 | .113 | Students who did not agree that they can develop friendships with classmates were more likely to be low distraction. |
| SCC | OLRQ 13 < Neutral | 126 | 0.238 | 0.560 | 0.945 | -0.057 | 0.792 | .192 | Students who did not agree that they can pay attention to other students social actions were more likely to be low distraction. |
| GS | SRLQ 1 = Strongly Agree | 131 | 0.247 | 0.548 | 0.925 | -0.082 | 0.716 | .060 | Students who strongly agreed that they set standards for their assignments in online courses were more likely to be low distraction. |
| GS | SRLQ 3 > Neutral | 241 | 0.455 | 0.554 | 0.935 | -0.167 | 0.374 | <.001 | Students who agreed that they keep a high standard for their learning in online courses were more likely to be low distraction. |
| GS | SRLQ 4 > Neutral | 238 | 0.449 | 0.561 | 0.947 | -0.127 | 0.505 | .004 | Students who agreed that they set goals to help manage their |

| | | | | | | | | |
|---|---|---|---|---|---|---|---|---|
| | | | | | | | | study time for online courses were more likely to be low distraction. |
| GS | SRLQ 5 > Neutral | 215 | 0.406 | 0.558 | 0.943 | -0.113 | 0.588 | .009 | Students who agreed that they do not compromise the quality of their work because it is online were more likely to be low distraction. |
| ES | SRLQ 9 > Neutral | 232 | 0.438 | 0.558 | 0.941 | -0.135 | 0.492 | .002 | Students who agreed that they choose a time with few distractions for studying for online courses were more likely to be low distraction. |
| TMTS | SRLQ 10 > Neutral | 124 | 0.234 | 0.556 | 0.939 | -0.063 | 0.771 | .146 | Students who agreed that they try to take more thorough notes for their online courses were more likely to be low distraction. |
| TMTS | SRLQ 16 > Neutral | 160 | 0.302 | 0.548 | 0.925 | -0.100 | 0.661 | .021 | Students who agreed that they still try to distribute their studying time evenly across the week were more likely to be low distraction. |
| HS | SRLQ 17 < Neutral | 117 | 0.221 | 0.544 | 0.919 | -0.081 | 0.715 | .062 | Students who did not agree that they find someone knowledgeable in course content to consult when they need help were more likely to be low distraction. |
| HS | SRLQ 18 < Neutral | 142 | 0.268 | 0.530 | 0.894 | -0.129 | 0.590 | .003 | Students who did not agree that they share their problems with classmates online so everyone knows what we are struggling |

| | |
|---|---|
| | with were more likely to be low distraction. |

*Note*.
LC: learner-content engagement; LI: learner-instructor engagement; SCC: social competencies with classmates; SCI: social competencies with instructors; TC: technical competencies; ES: environment structuring; GS: goal setting; HS: help-seeking; TMTS: time management and task strategies.

[a]: *p*-value of chi-square test.

*Step 2: Two Antecedent Attributes in ARM*

In the second step, 42 rules with two attributes were identified (see Appendix A). Seven were excluded due to redundancy. More than half of the rules (*n* = 24) combined an SRL strategy with attributes from the engagement strategy and OLR. Nearly one-third of the rules (*n* = 14) paired two SRL strategies. Less frequently observed were rules pairing two engagement strategies (*n* = 3), those combining an engagement strategy with OLR (*n* = 1), or those pairing two OLR attributes (*n* = 1). We found no rules that included student demographic characteristics and majors in this step. Overall, students with lower digital distraction were most frequently characterized by agreeing or more strongly agreeing with choosing a time with few distractions for studying, distributing studying time evenly across the week, and keeping a high standard for learning (60% of rules; SRLQ 3, 9, 16), as well as agreeing with statements related to learner-learner and learner-instructor engagement strategies (26% of rules; ESOQ 22, 19, and 28). Again, disagreement with peer help-seeking (SRLQ 18) emerged in multiple rules. Overall, SRL strategies (i.e., goal setting, environment structuring, and time management) emerged as the most prominent factor associated with lower digital distraction scores, followed by certain aspects of learner-content and learner-instructor engagement.

**Clustering of Rules**

The PAM cluster analysis classified the 42 identified rules from Step 2 into four clusters, each representing distinctive patterns of learning strategy combinations associated with lower distraction (see Figure 2). The clustering analysis showed that combinations of environment structuring, time management, learner content engagement, goal setting, and learner instructor engagement most frequently characterized students in the lower distraction group. Figure S1 in Appendix A provides a detailed overview of each rule's attributes and their respective cluster assignments. Each cluster was characterized as follows:

**Figure 2**

*Four Clusters of Rules Characterizing Digital Distraction Decrease*

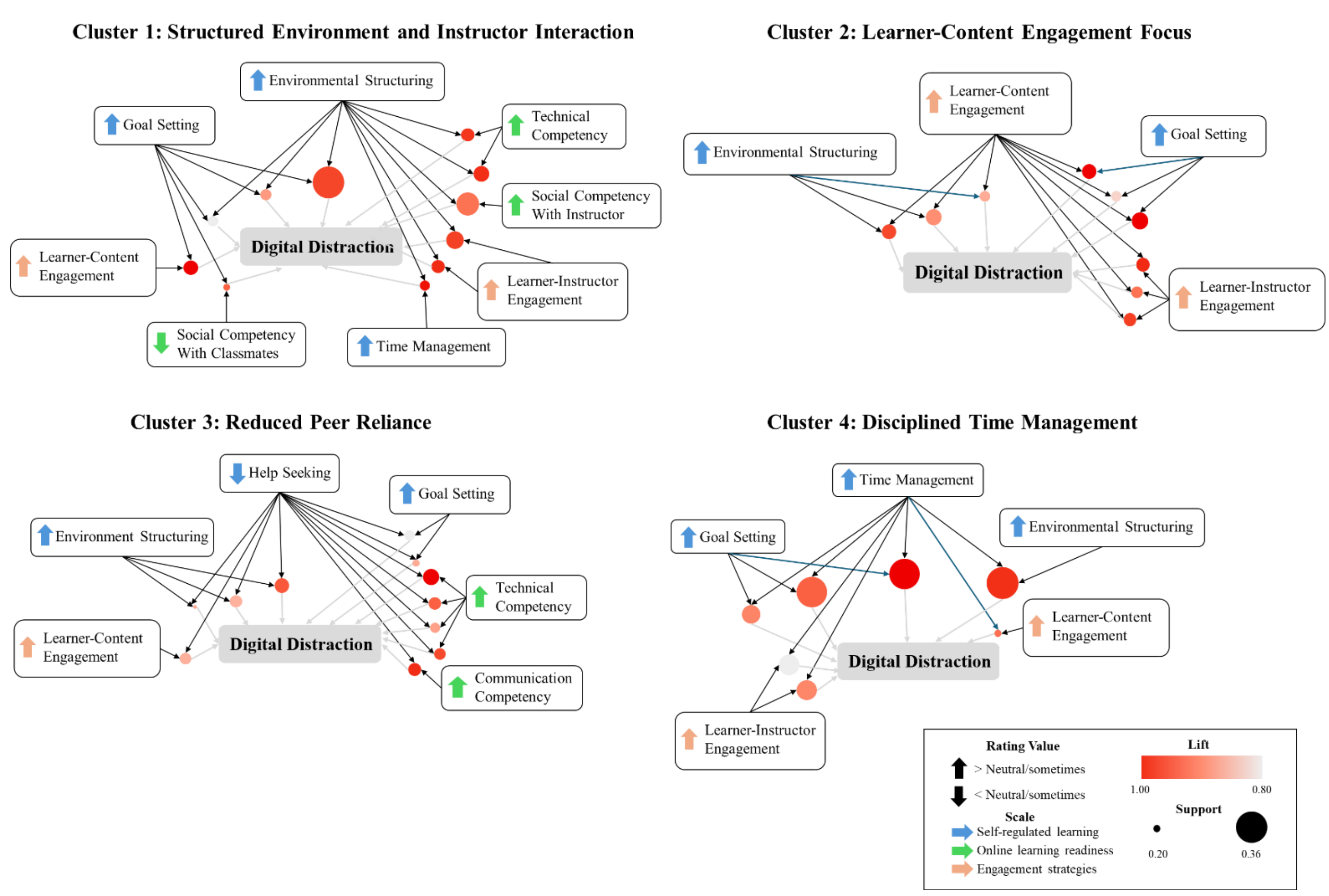

**Cluster 1: Structured Environment and Instructor Interaction**. This cluster comprised 12 rules that reflected strong co-occurrence of environment structuring and goal setting behaviors, often paired with learner-instructor engagement and technical competence. Students reported choosing specific times with minimal distractions for online studying (SRLQ 9) and maintaining high standards for their learning performance (SRLQ 3). These behaviors were consistently paired with learner-instructor engagement strategies, such as acknowledging the importance of short videos to enhance instructor presence (ESOQ 15). Students in this cluster also demonstrated confidence in their technical competencies (OLRQ 1) and ability to communicate effectively with instructors (OLRQ 11). Together, these findings suggested that students in this cluster mitigated distraction by using structured and goal-driven approaches, often engaging actively with instructors.

**Cluster 2: Learner-Content Engagement Focus**. This cluster contained 10 rules where students consistently emphasized the importance of learner-content engagement strategies. Specifically, students reported behaviors such as using online resources to explore topics deeply (ESOQ 22) and applying course materials to realistic scenarios (ESOQ 28). These rules were closely aligned with environment structuring, where students chose when and where to study to minimize distractions, and with goal setting behaviors related to establishing clear standards for learning performance. The findings for Cluster 2 highlighted a strong learner-content focus, underscored by minimizing digital distractions when students focus on interacting deeply with learning content while managing study environments and goals effectively.

**Cluster 3: Reduced Peer Reliance**. This cluster comprised 13 rules that uniquely featured lower reported peer-help-seeking behaviors combined with high environment structuring, goal setting, instructor engagement, and technical competencies. Unlike other clusters, students in this group reported a reduced reliance on peer collaboration for solving problems in online courses (SRLQ 18). However, these behaviors co-occurred with high agreement on environment structuring and goal setting strategies. Additionally, students in this cluster prioritized interactions with instructors (ESOQ 14, ESOQ 19). Another defining characteristic of Cluster 3 was the reported confidence in technical competencies (OLRQ 1–5).

**Cluster 4: Disciplined Time Management**. This cluster consisted of 7 rules that highlighted time management strategies as a defining feature. Students reported behaviors such as evenly distributing their study time across the week to maintain a steady learning pace (SRLQ 16). Time management strategies in this cluster co-occurred with strong agreement on environment structuring and goal setting behaviors, as well as the importance of both learner-content and learner-instructor engagement strategies. Thus, students in this cluster mitigated distraction through disciplined time management combined with structured study environments and active engagement strategies.

## Discussion

The objective of this study was to identify online learning strategies associated with the lower level of digital distraction among college students using unsupervised machine learning techniques. The findings highlighted the critical role of SRL strategies and emphasized the importance of engagement strategies and technical competencies within OLR in minimizing distractions. These insights offer valuable guidance for educators seeking to design interventions that enhance students' ability to manage distractions effectively. The following sections discuss these key findings in detail, explaining the contributions of SRL, engagement strategies, and OLR, followed by practical implications for educators.

### Self-regulated Learning

The study identified several SRL strategies that are significantly associated with the lower level of digital distraction in online learning environments. Among the strategies identified, goal setting, environment structuring, and time management emerged as significant predictors of lower distraction. These strategies enable learners to proactively create a structured approach to their studies, which is essential in the unstructured and often distraction-laden context of online learning environments. The findings aligned with the existing literature on SRL. Prior research has consistently demonstrated the importance of goal setting and environment structuring in enhancing attention and academic outcomes. For instance, Deng et al. (2024) highlighted that goal setting acted as a cognitive anchor, helping students prioritize essential tasks over distractions. Similarly, Chien et al. (2022) emphasized the importance of a well-structured learning environment, noting that students who actively managed their surroundings experienced fewer disruptions and achieved better learning outcomes. Khan and Awan (2017) mentioned that effective time management reduced the risk of Internet addiction and digital distractions by establishing clear time-use boundaries. As Wang et al. (2022) noted, effective time management can reduce the cognitive strain associated with multitasking by creating focused periods for academic work.

Interestingly, the study revealed a counterintuitive finding: a lower reliance on peer help-seeking was associated with lower digital distractions. While peer interactions are often regarded as supportive of engagement and academic achievement (e.g., Flanigan & Babchuk, 2022; Liao & Wu, 2022), the findings of this study suggested that students who reported lower distraction tended to report less frequent peer-focused help seeking. This phenomenon might stem from the dual nature of peer interactions in such settings, where the boundary between academic collaboration and non-academic socializing is often blurred (Ullah & Anwar, 2020). On the one hand, peer interactions serve as a vital resource for collaborative learning, enabling students to share resources and support each other in overcoming challenges (Wentzel & Watkins, 2002). However, informal interactions in online environments can also introduce opportunities for off-task behaviors. For instance, platforms designed for academic discussions, such as group chats or other messaging tools, may simultaneously serve as avenues for casual conversations, diverting attention from the primary academic objectives (Ullah & Anwar, 2020). Notifications, social media-like interfaces, and the absence of real-time supervision might further exacerbate this blurring of boundaries. Furthermore, this issue might be complicated by the emotional and cognitive states of learners. Wu (2017) indicated that when learners felt frustrated or bored with a course, they may turn to peer interactions as a form of escape.

**Engagement Strategies**

Our results revealed that learner-learner engagement strategies appeared less frequently among students who reported lower distraction compared with other engagement dimensions, such as engagement with content and instructors. The findings of this study align with Martin and Bolliger's (2018) investigation that engagement strategies supporting instructor interaction were valued more than others. Martin and Bolliger (2018) identified that clear instructions and guidelines given by instructors were deemed essential for effective engagement. Similarly, the current study found that timely feedback and interactive features are significantly associated with lower distractions.

In terms of learner-content engagement, Martin and Bolliger emphasized the importance of structured discussions and real-world applications, such as case studies and authentic projects, to deepen cognitive engagement. This study corroborates these findings, showing that authentic content also mitigated distractions by offering resources for exploring topics in more depth and realistic scenarios for students to apply content.

While items related to learner-learner engagement were rated as important by a majority in Martin and Bolliger's (2018) study, some students found collaborative work less valuable. Many participants in their investigation regarded peer discussion as busy work, lacking depth and meaningful interaction. Meanwhile, some participants complained synchronous sessions were viewed as social rather than educational. The current study aligns with this critique and has new explanations from a digital distraction perspective. We found that students with lower digital distractions placed less emphasis on peer interaction. Kyei-Blankson et al. (2019) similarly found that teaching presence and learner-instructor interaction were deemed most influential for online learning performance. They emphasized that instructor presence, which involves designing instructional materials, facilitating discourse, and providing direct instruction (Lin & Sun, 2024), was perceived as the most significant factor influencing student learning in online environments (Kyei-Blankson et al., 2019). These findings highlight the need for thoughtful design of peer collaborative activities. While learner-learner strategies may foster online learning, they may not play a significant role in online learning and could even contribute to perceptions of inefficiency if not thoughtfully designed (Marquez, 2023).

**Online Learning Readiness**

Regarding OLR, confidence in using computer technologies for learning was frequently observed among students who reported lower distraction. This suggests that technical competence tends to coexist with lower distraction reports. In Yu's (2018) measurement framework, the validation of the instrument underscored technical competency as a critical component of OLR, positively affecting academic achievement and learner satisfaction. Technical competency enables learners to utilize technology efficiently and provides more focus on their academic goals. From Rosen et al (2013)'s observation, students adept at employing technology for learning tasks were better equipped to resist distractions and maintain focus. Thus, institutions can consider integrating targeted training programs aimed at enhancing technical competencies and digital self-efficacy among online learners. By fostering robust technical readiness, educators can empower students to navigate online learning effectively, ultimately enhancing their academic success.

## Interconnectedness of Learning Strategies

In this context, self-regulated learning strategies, engagement strategies, and online learning readiness appeared together in patterned ways in our association rules. Across clusters, these constructs tended to co-appear among students who self-reported lower distraction. In the 42 paired rules, at least 67% of the rules emphasized SRL strategies such as environment structuring, goal setting, and time management as critical behaviors for significantly associating with lower levels of digital distraction.

In addition, the PAM clustering of rules mined four clusters and further demonstrated diverse combinations of these interrelated learning strategies. For instance, some students prioritized time management in maintaining steady progress as well as structured environments (see cluster 4), while others focused on the importance of learner-content engagement paired with the SRL strategies (see cluster 2). Cluster 3 was uniquely characterized by a low reliance on help-seeking from peers. These learners relied less on peer collaboration and more on their ability to structure environments, set goals, and engage directly with instructors. Despite these variations, all clusters underscored the pivotal role of SRL strategies, combined with engagement behaviors and individual competencies, in fostering effective online learning while reducing distractions. These clusters imply that students adopt diverse yet interconnected strategies to navigate online learning. By understanding these clusters, educators can design more personalized interventions to meet the needs of diverse learners in online environments.

## Implications for Educators

By examining the associated learning strategies, we bring a new perspective to the field of problematic use of the Internet and technology when students study online. The results of this study provide implications for educators by revealing how and which components to focus on to reduce the level of digital distractions in online learning environments.

SRL strategies, such as goal setting, environment structuring, and time management, are commonly present among students who report relatively low distraction in online courses. These strategies empower students to take ownership of their learning process and maintain focus. Educators can help students understand the consequences of their actions to raise awareness about the adverse effects of digital distractions, which may lead to more self-regulation (Flanigan et al., 2023). Besides, educators can incorporate explicit training on self-control skills within course structures. For example, embedding workshops or modules on creating structured study schedules and using online self-control tools like Pomodoro timers, task prioritization apps, and notification blockers can reinforce these practices (Biedermann et al., 2024; Lyngs et al., 2024).

Furthermore, interactive course designs, incorporating frequent instructor-led feedback and interaction opportunities, also tended to appear in the same profiles as lower distraction. Instructors can design synchronous sessions with engaging elements like real-time polls, Q&A sessions, and personalized communication to sustain attention.

Finally, the mixed results regarding learner-learner interactions suggest the need for careful integration of peer collaboration mechanisms. While peer interactions can be beneficial, improper design might contribute to distractions, particularly when these interactions are not goal-directed (Pozzi, 2010). Alternative support mechanisms, such as structured group activities, could be more effective in maintaining focus while fostering collaboration (Pozzi, 2010).

## Conclusions and Future Directions

This study researched an important yet underexplored topic of education: mitigating digital distraction in online learning environments. Through the innovative use of unsupervised machine learning techniques, including ARM and clustering analysis, this research examined associated learning strategies with the lower level of digital distractions. Key findings showed the SRL strategies, such as goal setting, environment structuring, and time management, were frequently reported together with lower digital distraction. Contrary to common assumptions, peer interaction strategies did not positively correlate with distraction reduction, highlighting the need for a deeper exploration of this field. By employing ARM and clustering analysis, this study extends the current literature by identifying co-occurring strategies rather than isolated effects.

Despite the contributions, this study is not without limitations. First, the sample, drawn from college students in a public university in southeastern America, may not represent the broader population. Second, self-reported surveys may introduce subjective biases, as participants' perceptions and recall may not accurately reflect their behaviors. However, this exploratory study provides a good start for future research on mitigating digital distractions, offering valuable preliminary insights and setting a direction for more comprehensive investigations.

Future research could complement self-reported data with objective measures of digital distraction, such as screen time tracking or behavioral observation. Longitudinal studies can also be conducted to examine the impact of SRL strategies on digital distractions over time. In addition, qualitative investigations could offer deeper insights into why peer interactions appear less effective in mitigating distractions in online learning environments.

**Availability of data and materials:** The data and code supporting the findings of this study are available at https://doi.org/10.17605/OSF.IO/P85S3

**Funding**: Not applicable.

**Acknowledgments**: Not applicable.

**Reference**

Abed, M. G., & Shackelford, T. K. (2024). Saudi parents' perspectives on the use of touch screen tablets for children with learning disabilities. *International Journal of Inclusive Education, 28*(8), 1324-1338. https://doi.org/10.1080/13603116.2021.1991491

Agrawal, P., Sahana, H. S., & De', R. (2017). Digital distraction. In R. Baguma, R. De', & T. Janowski (Eds.) *Proceedings of the 10th International Conference on Theory and Practice of Electronic Governance* (pp. 191-194). New Delhi, AA, India. https://doi.org/10.1145/3047273.304732

Agrawal, R., Imieliński, T., & Swami, A. (1993). Mining association rules between sets of items in large databases. In P. Buneman, & S. Jajodia (Eds.) *Proceedings of the 1993 ACM SIGMOD International Conference on Management of Data* (pp. 207-216). Washington D.C., USA. https://doi.org/10.1145/170035.170072

Agrawal, R., & Srikant, R. (1994). Fast algorithms for mining association rules. In *Proceedings of the 20th International Conference Very Large Data Bases (VLDB), 1215*, 487-499.

Aivaz, K. A., & Teodorescu, D. (2022). College students' distractions from learning caused by multitasking in online vs. face-to-face classes: A case study at a public university in Romania. *International Journal of Environmental Research and Public Health, 19*(18), 11188. https://doi.org/10.3390/ijerph191811188

Anderson, J. C., & Gerbing, D. W. (1988). Structural equation modeling in practice: A review and recommended two-step approach. *Psychological Bulletin, 103*, 411–423. https://doi.org/10.1037/0033-2909.103.3.411

Aristovnik, A., Karampelas, K., Umek, L., & Ravšelj, D. (2023). Impact of the COVID-19 pandemic on online learning in higher education: a bibliometric analysis. *Frontiers in Education, 8*, 1-13. https://doi.org/10.3389/feduc.2023.1225834

Barnard, L., Lan, W. Y., To, Y. M., Paton, V. O., & Lai, S. L. (2009). Measuring self-regulation in online and blended learning environments. *The Internet and Higher Education, 12*(1), 1-6. https://doi.org/10.1016/j.iheduc.2008.10.005

Baturay, M. H., & Toker, S. (2015). An investigation of the impact of demographics on cyberloafing from an educational setting angle. *Computers in Human Behavior, 50*, 358-366. https://doi.org/10.1016/j.chb.2015.03.081

Biedermann, D., Kister, S., Breitwieser, J., Weidlich, J., & Drachsler, H. (2024). Use of digital self-control tools in higher education–a survey study. *Education and Information Technologies, 29*(8), 9645-9666. https://doi.org/10.1007/s10639-023-12198-2

Bradley, V. M. (2021). Learning Management System (LMS) use with online instruction. *International Journal of Technology in Education, 4*(1), 68–92. https://doi.org/10.46328/ijte.36

Brock, G., Pihur, V., Datta, S., & Datta, S. (2008). clValid: An R package for cluster validation. *Journal of Statistical Software, 25*(4), 1–22. https://doi.org/10.18637/jss.v025.i04

Campbell, M., Edwards, E. J., Pennell, D., Poed, S., Lister, V., Gillett-Swan, J., ... & Nguyen, T. A. (2024). Evidence for and against banning mobile phones in schools: A scoping review. *Journal of Psychologists and Counsellors in Schools, 34*(3), 242–265. https://doi.org/10.1177/20556365241270394

Chen, L., Nath, R., & Insley, R. (2014). Determinants of digital distraction: A cross-cultural investigation of users in Africa, China and the US. *Journal of International Technology and Information Management, 23*(3), 145–172. https://doi.org/10.58729/1941-6679.1080

Chen, Q., Yan, Z., Moeyaert, M., & Bangert-Drowns, R. (2025). Mobile multitasking in learning: A meta-analysis of effects of mobilephone distraction on young adults' immediate recall. *Computers In Human Behavior, 162*, 108432. https://doi.org/10.1016/j.chb.2024.108432

Chien, H. Y., Yeh, Y. C., & Kwok, O. M. (2022). How Online Learning Readiness Can Predict Online Learning Emotional States and Expected Academic Outcomes: Testing a Theoretically Based Mediation Model. *Online Learning, 26*(4), 193–208. https://doi.org/10.24059/olj.v26i4.3483

Crawford, J., Butler-Henderson, K., Rudolph, J., Malkawi, B., Glowatz, M., Burton, R., ... & Lam, S. (2020). COVID-19: 20 countries' higher education intra-period digital pedagogy responses. *Journal of Applied Learning & Teaching, 3*(1), 1–20. https://doi.org/10.37074/jalt.2020.3.1.7

Dai, Y., Luan, L., & Lin, X. (2023). The effects of online learning readiness on self-regulated learning for the first-time online learning students. *Asian Journal of Distance Education, 18*(2), 42-62.

Dang, L., Kwan, L. Y. Y., Zhang, M. X., & Wu, A. M. (2024). Cognitive and affective correlates of cyber-slacking in Chinese university students. *The Asia-Pacific Education Researcher, 33*(3), 545–557. https://doi.org/10.1007/s40299-023-00752-y

Deng, L. (2020). Laptops and mobile phones at self-study time: Examining the mechanism behind interruption and multitasking. *Australasian Journal of Educational Technology, 36*(1), 55-67. https://doi.org/10.14742/ajet.5048

Deng, L., Zhou, Y., & Broadbent, J. (2024). Distraction, multitasking and self-regulation inside university classroom. *Education and Information Technologies, 29*, 23957–23979. https://doi.org/10.1007/s10639-024-12786-w

Flanigan, A. E., & Babchuk, W. A. (2022). Digital distraction in the classroom: Exploring instructor perceptions and reactions. *Teaching in Higher Education, 27*(3), 352-370. https://doi.org/10.1080/13562517.2020.1724937

Flanigan, A. E., Brady, A. C., Dai, Y., & Ray, E. (2023). Managing student digital distraction in the college classroom: A self-determination theory perspective. *Educational Psychology Review, 35*(2), 60. https://doi.org/10.1007/s10648-023-09780-y

Forero, C. G., Maydeu-Olivares, A., & Gallardo-Pujol, D. (2009). Factor analysis with ordinal indicators: A Monte Carlo study comparing DWLS and ULS estimation. *Structural Equation Modeling, 16*(4), 625-641. https://doi.org/10.1080/10705510903203573

Gerow, J. E., Galluch, P. S., & Thatcher, J. B. (2010). To slack or not to slack: Internet usage in the classroom. *Journal of Information Technology Theory and Application, 11*(3), 5–23.

Getenet, S., Cantle, R., Redmond, P., & Albion, P. (2024). Students' digital technology attitude, literacy and self-efficacy and their effect on online learning engagement. *International Journal of Educational Technology in Higher Education, 21*(1), 3. https://doi.org/10.1186/s41239-023-00437-y

Göl, B., Özbek, U., & Horzum, M. B. (2023). Digital distraction levels of university students in emergency remote teaching. *Education and Information Technologies, 28*(7), 9149-9170. https://doi.org/10.1007/s10639-022-11570-y

Gupta, G. K., Strehl, A., & Ghosh, J. (1999). Distance based clustering of association rules. In *Proceedings of Artificial Neural Networks in Engineering, 9*(1999), 759–764.

Hahsler, M., Grün, B., & Hornik, K. (2005). arules-A computational environment for mining association rules and frequent item sets. *Journal of Statistical Software, 14*(15), 1–25. https://doi.org/10.18637/jss.v014.i15

Hahsler, M., & Chelluboina, S. (2011). Visualizing association rules: Introduction to the R-extension package arulesViz. *R project module, 6*, 223-238.

Haque, U. M., Kabir, E., & Khanam, R. (2024). Investigating school absenteeism and refusal among Australian children and adolescents using Apriori association rule mining. *Scientific Reports, 14*(1), 1907. https://doi.org/10.1038/s41598-024-51230-4

Hatlevik, O. E., & Bjarnø, V. (2021). Examining the relationship between resilience to digital distractions, ICT self-efficacy, motivation, approaches to studying, and time spent on individual studies. *Teaching and Teacher Education, 102*, 103326. https://doi.org/10.1016/j.tate.2021.103326

Hu, L. T., & Bentler, P. M. (1999). Cutoff criteria for fit indexes in covariance structure analysis: Conventional criteria versus new alternatives. *Structural Equation Modeling: A Multidisciplinary Journal, 6*(1), 1-55. https://doi.org/10.1080/10705519909540118

Iacobucci, D., Posavac, S. S., Kardes, F. R., Schneider, M. J., & Popovich, D. L. (2015). The median split: Robust, refined, and revived. *Journal of Consumer Psychology, 25*(4), 690–704. https://doi.org/10.1016/j.jcps.2015.06.014

Kingir, S., Gok, B., & Bozkir, A. S. (2020). Exploring Relations among Pre-Service Science Teachers' Motivational Beliefs, Learning Strategies and Constructivist Learning Environment Perceptions through Unsupervised Data Mining. *Journal of Baltic Science Education, 19*(5), 804-823. https://doi.org/10.33225/jbse/20.19.804

Khan, A., Egbue, O., Palkie, B., & Madden, J. (2017). Active learning: Engaging students to maximize learning in an online course. *Electronic Journal of E-learning, 15*(2), 107-115.

Khan, H. U., & Awan, M. A. (2017). Possible factors affecting internet addiction: a case study of higher education students of Qatar. *International Journal of Business Information Systems, 26*(2), 261-276. https://doi.org/10.1504/IJBIS.2017.086339

Kraushaar, J. M., & Novak, D. C. (2010). Examining the affects of student multitasking with laptops during the lecture. *Journal of Information Systems Education, 21*(2), 241-252.

Kumar, C., Rangappa, K. B., Suchitra, S., & Gowda, H. (2024). Digital distractions during blended learning and its negative repercussions: an empirical analysis. *Asian Association of Open Universities Journal, 19*(1), 1-18. https://doi.org/10.1108/AAOUJ-02-2023-0024

Kyei-Blankson, L., Ntuli, E., & Donnelly, H. (2019). Establishing the importance of interaction and presence to student learning in online environments. *Journal of Interactive Learning Research, 30*(4), 539-560.

Lin, L., King, R. B., Fu, L., & Leung, S. O. (2024). Information and communication technology engagement and digital reading: How meta‐cognitive strategies impact their relationship. *British Journal of Educational Technology, 55*(1), 277-296. https://doi.org/10.1111/bjet.13355

Lin, X., & Dai, Y. (2022). An exploratory study of the effect of online learning readiness on self-regulated learning. *International Journal of Chinese Education, 11*(2), https://doi.org/10.1177/2212585X221111938

Lin, X., & Sun, Q. (2023). Student‐generated memes as a way to facilitate online discussion for adult learners. *Psychology in the Schools, 60*(12), 4826-4840. https://doi.org/10.1002/pits.22884

Lin, X., & Sun, Q. (2024). Discussion activities in asynchronous online learning: Motivating adult learners' interactions. *The Journal of Continuing Higher Education, 72*(1), 84-103. https://doi.org/10.1080/07377363.2022.2119803

Lin, X., Sun, Q., & Zhang, X. (2024). Increasing student online interactions: Applying the video timeline-anchored comment (VTC) tool to asynchronous online video discussions. *International Journal of Human–Computer Interaction, 40*(19), 5910-5922. https://doi.org/10.1080/10447318.2023.2247554

Lund Dean, K., & Jolly, J. P. (2012). Student identity, disengagement, and learning. *Academy of Management Learning & Education, 11*(2), 228-243. https://doi.org/10.5465/amle.2009.0081

Lyngs, U., Lukoff, K., Slovak, P., Inzlicht, M., Freed, M., Andrews, H., ... & Shadbolt, N. (2024). "I finally felt I had the tools to control these urges": Empowering Students to Achieve Their Device Use Goals With the Reduce Digital Distraction Workshop. In F. F. Mueller, P. Kyburz, J. R. Williamson, C. Sas, M. L. Wilson, P. T. Dugas, & I. Shklovski (Eds.). P*roceedings of the CHI Conference on Human Factors in Computing Systems* (pp. 1-23). Honolulu, HI, USA. https://doi.org/10.1145/3613904.364294

Maechler, M., Rousseeuw, P., Struyf, A., Hubert, M., & Hornik, K. (2014). *cluster: Cluster Analysis Basics and Extensions* (Version 2.0. 7–1). R package. https://rdrr.io/cran/parameters/src/R/cluster_analysis.R

Marquez, N. (2023). *Influence of Applications and Usage of Technology on Student Engagement in the Classroom a Qualitative Study* (Publication No. 30425084.) [Doctoral dissertation, Liberty University]. ProQuest Dissertations and Theses Global.

Martin, F., & Bolliger, D. U. (2018). Engagement matters: Student perceptions on the importance of engagement strategies in the online learning environment. *Online Learning, 22*(1), 205-222. https://doi.org/10.24059/olj.v22i1.1092

Martin, F., Long, S., Haywood, K., & Xie, K. (2025). Digital distractions in education: a systematic review of research on causes, consequences and prevention strategies. *Educational Technology Research and Development*, 1-29. https://doi.org/10.1007/s11423-025-10550-6

Masry-Herzallah, A., & Watted, A. (2024). Technological self-efficacy and mindfulness ability: Key drivers for effective online learning in higher education beyond the COVID-19 era. *Contemporary Educational Technology, 16*(2), ep505. https://doi.org/10.30935/cedtech/14336

Moore, M. J. (1993). *Three types of interaction*. In K. Harry, M. John, & D. Keegan (Eds.), Distance education theory (pp. 19–24). Routledge.

Muir, T., Wang, I., Trimble, A., Mainsbridge, C., & Douglas, T. (2022). Using interactive online pedagogical approaches to promote student engagement. *Education Sciences, 12*(6), 415. https://doi.org/10.3390/educsci12060415

Nabung, A. (2024). The impact of multitasking with digital devices on classroom learning: A critical review on the future of digital distraction in education. *US-China Education Review, 14*(6), 369-383. https://doi.org/10.17265/2161-623X/2024.06.005

Pintrich, P. R. (2004). A conceptual framework for assessing motivation and self-regulated learning in college students. *Educational Psychology Review, 16,* 385-407. https://doi.org/10.1007/s10648-004-0006-x

Pozzi, F. (2010). Using Jigsaw and Case Study for supporting online collaborative learning. *Computers & Education, 55*(1), 67-75. https://doi.org/10.1016/j.compedu.2009.12.003

Rivers, D. J., Nakamura, M., & Vallance, M. (2022). Online self-regulated learning and achievement in the era of change. *Journal of Educational Computing Research, 60*(1), 104-131. https://doi.org/10.1177/07356331211025108

Rousseeuw, P. J. (1987). Silhouettes: A graphical aid to the interpretation and validation of cluster analysis. *Journal of Computational and Applied Mathematics, 20*, 53-65. https://doi.org/10.1016/0377-0427(87)90125-7

Rosen, L. D. (2017). The distracted student mind—enhancing its focus and attention. *Phi Delta Kappan, 99*(2), 8-14. https://doi.org/10.1177/0031721717734183

Rosen, L. D., Carrier, L. M., & Cheever, N. A. (2013). Facebook and texting made me do it: Media-induced task-switching while studying. *Computers in Human Behavior, 29*(3), 948-958. https://doi.org/10.1016/j.chb.2012.12.001

Sahdan, S., Masek, A., & Zainal Abidin, N. A. (2017). Student's readiness on self-regulated learning implementation for 21st century learning approaches. *Pertanika Journal of Social Sciences & Humanities, 25*(S), 195-204.

Samson, P. J. (2010). Deliberate engagement of laptops in large lecture classes to improve attentiveness and engagement. *Computers in Education, 20*(2), 22-37.

Shi, H., Zhou, Y., Dennen, V. P., & Hur, J. (2024). From unsuccessful to successful learning: profiling behavior patterns and student clusters in massive open online courses. *Education and Information Technologies, 29*(5), 5509–5540. https://doi.org/10.1007/s10639-023-12010-1

Sun, W., & Chao, M. (2024). Exploring the influence of excessive social media use on academic performance through media multitasking and attention problems: a three-dimension usage perspective. *Education and Information Technologies, 29*(18), 23981–24003. https://doi.org/10.1007/s10639-024-12811-y

Taghavi-Nejad, F. S., Fallah, N., & Lotfi Gaskaree, B. (2024). Mindfulness and Procrastination Among University EFL Learners: The Role of Attention Control and Self-Regulated Learning. *Psychological Reports*. https://doi.org/10.1177/00332941241287423

Taneja, A., Fiore, V., & Fischer, B. (2015). Cyber-slacking in the classroom: Potential for digital distraction in the new age. *Computers & Education, 82,* 141-151. https://doi.org/10.1016/j.compedu.2014.11.009

Ullah, A., & Anwar, S. (2020). The effective use of information technology and interactive activities to improve learner engagement. *Education Sciences, 10*(12), 349. https://doi.org/10.3390/educsci10120349

Wang, C. H., Salisbury-Glennon, J. D., Dai, Y., Lee, S., & Dong, J. (2022). Empowering college students to decrease digital distraction through the use of self-regulated learning strategies. *Contemporary Educational Technology, 14*(4), ep388. https://doi.org/10.30935/cedtech/12456

Wentzel, K. R., & Watkins, D. E. (2002). Peer relationships and collaborative learning as contexts for academic enablers. *School Psychology Review, 31*(3), 366-377. https://doi.org/10.1080/02796015.2002.12086161

Wu, J. Y. (2017). The indirect relationship of media multitasking self-efficacy on learning performance within the personal learning environment: Implications from the mechanism of perceived attention problems and self-regulation strategies. *Computers & Education, 106*, 56-72. https://doi.org/10.1016/j.compedu.2016.10.010

Wu, J. Y., & Xie, C. (2018). Using time pressure and note-taking to prevent digital distraction behavior and enhance online search performance: Perspectives from the load theory of attention and cognitive control. *Computers in Human Behavior, 88*, 244-254. https://doi.org/10.1016/j.chb.2018.07.008

Xia, Y., & Yang, Y. (2019). RMSEA, CFI, and TLI in structural equation modeling with ordered categorical data: The story they tell depends on estimation methods. *Behavior Research Methods, 51,* 409-428. https://doi.org/10.3758/s13428-018-1055-2

Yan, X., Zhang, C., & Zhang, S. (2009). Confidence metrics for association rule mining. *Applied Artificial Intelligence, 23*(8), 713-737. https://doi.org/10.1080/08839510903208062

Yu, T. (2018). Examining Construct Validity of the Student Online Learning Readiness (SOLR) Instrument Using Confirmatory Factor Analysis. *Online learning, 22*(4), 277-288. https://doi.org/10.24059/olj.v22i4.1297

Yu-Lin, H. O., Chien, C. H. O. U., & Chen-Hsuan, L. I. A. O. (2022). Using Unsupervised Machine Learning to Model Taiwanese High-School Students' Digital Distraction Profiles Concerning Internet Gaming Disorder. *International Conference on Computers in Education*. Retrieved from https://library.apsce.net/index.php/ICCE/article/view/4455

Zhao, Y. (2012). Association Rules. In Y. Zhao (Eds.), *R and data mining: Examples and case studies* (pp. 89-100). Academic Press.

Zimmerman, B. J. (2000). Attaining self-regulation. In M. Boekaerts, P. R. Pintrich, & M. Zeidner (Eds.), *Handbook of self-regulation* (pp. 13–39). Academic Press. https://doi.org/10.1016/B978-012109890-2/50031-7

## Appendix A

**Table S1**

*Associated Rules with Two Antecedent Attributes for Digital Distraction Decrease*

| Subscale | Attribute 1 | Attribute 2 | N | Supp. | Conf. | Lift | Phi | Odds Ratio |
|---|---|---|---|---|---|---|---|---|
| LC + SCI | ESOQ 22=Agree | OLRQ 10>Neutral | 106 | 0.200 | 0.502 | 0.848 | -0.149 | 0.539 |
| LC + ES | ESOQ 22=Agree | SRLQ 6>Neutral | 122 | 0.230 | 0.528 | 0.891 | -0.115 | 0.624 |
| LC + ES | ESOQ 22=Agree | SRLQ 8>Neutral | 127 | 0.240 | 0.518 | 0.875 | -0.140 | 0.564 |
| LC + ES | ESOQ 22=Agree | SRLQ 9>Neutral | 113 | 0.213 | 0.514 | 0.867 | -0.135 | 0.573 |
| LC + GS | ESOQ 22=Agree | SRLQ 1>Neutral | 130 | 0.245 | 0.533 | 0.899 | -0.112 | 0.632 |
| LC + GS | ESOQ 22=Agree | SRLQ 2>Neutral | 122 | 0.230 | 0.533 | 0.899 | -0.106 | 0.647 |
| LC + GS | ESOQ 22=Agree | SRLQ 4>Neutral | 112 | 0.211 | 0.507 | 0.855 | -0.147 | 0.544 |
| LC + GS | ESOQ 27=Agree | SRLQ 3>Neutral | 114 | 0.215 | 0.533 | 0.899 | -0.100 | 0.661 |
| LC + LI | ESOQ 28=Agree | ESOQ 15>Neutral | 115 | 0.217 | 0.523 | 0.882 | -0.120 | 0.611 |
| LC + LI | ESOQ 28=Agree | ESOQ 19>Neutral | 120 | 0.226 | 0.531 | 0.896 | -0.108 | 0.642 |
| LC + TSTM | ESOQ 22>Neutral | SRLQ 16>Neutral | 117 | 0.221 | 0.527 | 0.890 | -0.113 | 0.628 |
| LC + HS | ESOQ 28>Neutral | SRLQ 18<Neutral | 117 | 0.221 | 0.511 | 0.862 | -0.145 | 0.551 |
| LI + LC | ESOQ 19=Strongly Agree | ESOQ 22=Agree | 118 | 0.223 | 0.529 | 0.893 | -0.110 | 0.636 |
| LI + HS | ESOQ 19=Strongly Agree | SRLQ 18<Neutral | 119 | 0.225 | 0.531 | 0.897 | -0.107 | 0.645 |
| LI + HS | ESOQ 14>Neutral | SRLQ 18<Neutral | 118 | 0.223 | 0.532 | 0.897 | -0.105 | 0.648 |
| LI + TSTM | ESOQ 14>Neutral | SRLQ 16>Neutral | 126 | 0.238 | 0.516 | 0.872 | -0.143 | 0.557 |
| LI + TSTM | ESOQ 17>Neutral | SRLQ 16>Neutral | 125 | 0.236 | 0.525 | 0.887 | -0.124 | 0.603 |
| CC + HS | OLRQ 19>Neutral | SRLQ 18<Neutral | 115 | 0.217 | 0.520 | 0.878 | -0.124 | 0.600 |
| SCC + GS | OLRQ 12<Neutral | SRLQ 4>Neutral | 106 | 0.200 | 0.527 | 0.890 | -0.104 | 0.649 |
| SCI + ES | OLRQ 11>Neutral | SRLQ 9>Neutral | 143 | 0.270 | 0.526 | 0.887 | -0.139 | 0.564 |
| TC + SCI | OLRQ 1=Agree | OLRQ 10>Neutral | 106 | 0.200 | 0.533 | 0.899 | -0.094 | 0.674 |
| TC + ES | OLRQ 1=Agree | SRLQ 9>Neutral | 112 | 0.211 | 0.531 | 0.896 | -0.102 | 0.655 |
| TC + GS | OLRQ 1=Agree | SRLQ 3>Neutral | 119 | 0.225 | 0.531 | 0.897 | -0.107 | 0.645 |
| TC + HS | OLRQ 1>Neutral | SRLQ 18<Neutral | 114 | 0.215 | 0.514 | 0.867 | -0.136 | 0.570 |
| TC + HS | OLRQ 3>Neutral | SRLQ 18<Neutral | 134 | 0.253 | 0.532 | 0.898 | -0.118 | 0.618 |
| TC + HS | OLRQ 4>Neutral | SRLQ 18<Neutral | 118 | 0.223 | 0.527 | 0.889 | -0.114 | 0.625 |
| TC + HS | OLRQ 5>Neutral | SRLQ 18<Neutral | 121 | 0.228 | 0.524 | 0.884 | -0.123 | 0.604 |
| ES + LI | SRLQ 8=Strongly Agree | ESOQ 16>Neutral | 123 | 0.232 | 0.530 | 0.895 | -0.112 | 0.632 |
| ES + LI | SRLQ 9=Strongly Agree | ESOQ 15>Neutral | 110 | 0.208 | 0.531 | 0.897 | -0.099 | 0.662 |
| ES + GS | SRLQ 9=Strongly Agree | SRLQ 3>Neutral | 107 | 0.202 | 0.512 | 0.864 | -0.132 | 0.578 |
| ES + HS | SRLQ 6>Neutral | SRLQ 18<Neutral | 106 | 0.200 | 0.512 | 0.864 | -0.131 | 0.580 |
| ES + HS | SRLQ 7>Neutral | SRLQ 18<Neutral | 127 | 0.240 | 0.525 | 0.886 | -0.126 | 0.596 |
| ES + HS | SRLQ 8>Neutral | SRLQ 18<Neutral | 119 | 0.225 | 0.511 | 0.862 | -0.147 | 0.546 |
| ES + TSTM | SRLQ 8>Neutral | SRLQ 16>Neutral | 146 | 0.275 | 0.531 | 0.896 | -0.130 | 0.586 |
| ES + TSTM | SRLQ 9>Neutral | SRLQ 10>Neutral | 107 | 0.202 | 0.532 | 0.899 | -0.096 | 0.671 |
| GS + ES | SRLQ 1=Strongly Agree | SRLQ 9>Neutral | 108 | 0.204 | 0.522 | 0.881 | -0.115 | 0.620 |
| GS + HS | SRLQ 1>Neutral | SRLQ 18<Neutral | 113 | 0.213 | 0.500 | 0.844 | -0.162 | 0.512 |
| GS + HS | SRLQ 2>Neutral | SRLQ 18<Neutral | 108 | 0.204 | 0.512 | 0.864 | -0.133 | 0.575 |
| GS + ES | SRLQ 3>Neutral | SRLQ 9>Neutral | 196 | 0.370 | 0.530 | 0.894 | -0.194 | 0.401 |
| GS + TSTM | SRLQ 3>Neutral | SRLQ 16>Neutral | 141 | 0.266 | 0.528 | 0.891 | -0.132 | 0.582 |
| GS + TSTM | SRLQ 4>Neutral | SRLQ 16>Neutral | 141 | 0.266 | 0.532 | 0.898 | -0.123 | 0.605 |
| GS + TSTM | SRLQ 5>Neutral | SRLQ 16>Neutral | 124 | 0.234 | 0.525 | 0.887 | -0.122 | 0.606 |

*Note*.
LC: learner-content engagement; LI: learner-instructor engagement; SCC: social competencies with classmates; SCI: social competencies with instructors; TC: technical competencies; ES: environment structuring; GS: goal setting; HS: help-seeking; TMTS: time management and task strategies.

ESOQ questions: 14-‘The instructor creates a course orientation for students.’; 15-‘The instructor posts a “due date checklist” at the end of each instructional unit’; 16-‘The instructor posts grading rubrics for all assignments.’; 17-‘The instructor provides feedback using various modalities (e.g., text, audio, video, and visuals)’; 18-‘The instructor provides students with an opportunity to reflect’; 19-‘The instructor posts grading rubrics for all assignments’; 20-‘The instructor uses various features in synchronous sessions to interact with students’; 22-‘Students use optional online resources to explore topics in more depth.’; 27-‘Students have an opportunity to reflect on important elements of the course’; 28-‘Students work on realistic scenarios to apply content’.

OLRQ questions: 1-‘ I have a sense of self-confidence in using computer technologies for specific tasks’; 3-‘I feel comfortable using computers’; 4-‘I can explain the benefits of using computer technologies in learning’; 5-‘I am competent at integrating computer technologies into my learning activities’; 10-‘I am confident that I can timely inform the instructor when unexpected situations arise’; 11-‘I am confident that I can express my opinions to instructor respectfully’; 12-‘I am confident that I can develop friendships with my classmates'; 13-'I am confident that I can pay attention to other students' social actions'; 19-'I am able to express my opinion in writing so that others understand what I mean'.

SRLQ questions: 1-'I set standards for my assignments in online courses'; 2-'I set short-term(daily or weekly) goals as well as long-term goals (monthly or for the semester)'; 3-'I keep a high standard for my learning in my online courses'; 4-'I set goals to help me manage studying time for my online courses'; 5-'I don't compromise the quality of my work because it is online'; 6-'I choose the location where I study to avoid too much distraction'; 7-'I find a comfortable place to study'; 8-'I know where I can study most efficiently for online courses'; 9-'I choose a time with few distractions for studying for my online courses'; 10-'I try to take more thorough notes for my online courses'; 16-'Although we don't have to attend daily classes, I still try to distribute my studying time evenly across'; 17-'I find someone who is knowledgeable in course content so that I can consult with him or her when I need help'; 18-'I share my problems with my classmates online so we know what we are struggling with and how to solve our problems'.

**Figure S1**

*Associated Rules with Digital Distraction Decrease by Cluster*

| | | | Engagement Strategy Questionnaire | | | Online Self-Regulated Learning Questionnaire | | | | | | | | | |
|---|---|---|---|---|---|---|---|---|---|---|---|---|---|---|---|
| | | | LI | | | ES | | GS | | | | | HS | TSTM | |
| | | | 15 >N | 16 >N | 19 >N | 6 >N | 8 >N | 9 >N | 1 >N | 2 >N | 3 >N | 4 >N | 18 <N | 16 >N | 10 >N |
| Engagement Strategy Questionnaire | LC | 22 >N | | | 2 | 2 | 2 | 2 | 2 | 2 | | 2 | | 4 | |
| | | 27 =A | | | | | | | | | 1 | | | | |
| | | 28 >N | 2 | | 2 | | | | | | | | 3 | | |
| | | 14 >N | | | | | | | | | | | 3 | 4 | |
| | LI | 15 >N | | | | | | 1 | | | | | | | |
| | | 16 >N | | | | | 1 | | | | | | | | |
| | | 17 >N | | | | | | | | | | | | 4 | |
| | | 19 =SA | | | | | | | | | | | 3 | | |
| Online Self-Regulated Learning Questionnaire | ES | 6 >N | | | | | | | | | | | 3 | | |
| | | 7 >N | | | | | | | | | | | 3 | | |
| | | 8 >N | | | | | | | | | | | 3 | 4 | |
| | | 9 >N | | | | | | | | | 1 | | | | 1 |
| | GS | 1 >N | | | | | | 1 | | | | | 3 | | |
| | | 2 >N | | | | | | | | | | | 3 | | |
| | | 3 >N | | | | | | 1 | | | | | | 4 | |
| | | 4 >N | | | | | | | | | | | | 4 | |
| | | 5 >N | | | | | | | | | | | | 4 | |
| | HS | 18 <N | | | | | | | | | | | | | |
| Online Learning Readines Questionnaire | CC | 19 >N | | | | | | | | | | | 3 | | |
| | SCC | 12 <N | | | | | | | | | | 1 | | | |
| | SCI | 11 >N | | | | | | 1 | | | | | | | |
| | TC | 1 >N | | | | | | 1 | | | 1 | | 3 | | |
| | | 3 >N | | | | | | | | | | | 3 | | |
| | | 4 >N | | | | | | | | | | | 3 | | |
| | | 5 >N | | | | | | | | | | | 3 | | |

A: Agree; SA: Strongly Agree; N: Neutral; LC: learner-content engagement; LI: learner-instructor engagement; CC: communication competencies; SCC: social competencies with classmates; SCI: social competencies with instructors; TC: technical competencies; ES: environment structuring; GS: goal setting; HS: help-seeking; TMTS: time management and task strategies.

**Supplemental Document**

**Supplementary analysis**

We conducted a supplementary sensitivity analysis that focused specifically on the most extreme ends of the distribution of digital distraction. This analysis evaluates whether the core learning strategies associated with lower distraction in the main analysis remain stable when we restrict attention to only the lowest distraction students versus the highest distraction students, rather than using the full sample and a single median cut. This procedure allows us to assess the robustness of our main findings to alternative operationalizations of the outcome variable.

First, we extracted the bottom 25% (lowest distraction) to the top 25% (highest distraction) dataset. This resulted in two non-overlapping groups: 1) the lowest distraction group (bottom 25% of the digital distraction distribution): 174 participants, and 2) the highest distraction group (top 25% of the digital distraction distribution): 150 participants. Second, we ran association rule mining using only these two extreme groups. In this sensitivity association rule mining, participants in the lowest distraction quartile served as the consequent (outcome) of interest. That is, rules can be interpreted as significant learning strategies that are associated with the lowest level of digital distraction.

We preserved the original association rule mining evaluation metrics, including support, confidence, lift, phi coefficient, odds ratio, and chi-square tests of disproportionate appearance. These metrics quantify the strength and direction of the association between a given antecedent attribute and belonging to the lowest distraction level. Because quartile-based groups are smaller than the two groups created by the median split, we made one adjustment to the rule mining thresholds. Specifically, we reduced the minimum support threshold from 0.2 to 0.1 to reflect the smaller subgroup size and to allow frequent-enough itemsets to be discovered within a more selective subset of students. Other thresholds and filters were left unchanged.

Table 1 presents the significant single-antecedent rules for predicting participants in the lowest distraction quartile. The results of the sensitivity analysis largely replicated the main results. The association rule mining procedure again identified core self-regulated learning strategies, learner-content engagement, learner-instructor engagement strategies, and technical confidence as significant antecedents of belonging to the lowest distraction group. All of these antecedent attributes were already reported in the main manuscript and, therefore, are not newly introduced constructs. In other words, the quartile-based analysis did not reveal any qualitatively new predictors and did not elevate demographic variables such as gender, age, race, or major into significant rules, consistent with the primary analysis.

**Table 1**

*Significant Rules with Single Antecedent Attributes in Top and Bottom Quartiles*

| Subscale | Attribute | N | Supp. | Conf. | Lift | Phi | Odds Ratio | *p*[a] |
|---|---|---|---|---|---|---|---|---|
| LC | ESOQ 22 = Agree | 152 | 0.287 | 0.545 | 0.920 | -0.102 | 0.658 | .019 |
| TC | OLRQ 1 = Agree | 150 | 0.283 | 0.558 | 0.941 | -0.072 | 0.746 | .098 |
| SCI | OLRQ 10 > Neutral | 223 | 0.421 | 0.560 | 0.946 | -0.114 | 0.574 | 009 |
| SCI | OLRQ 11 > Neutral | 181 | 0.342 | 0.562 | 0.949 | -0.077 | 0.724 | .077 |
| GS | SRLQ 3 > Neutral | 241 | 0.455 | 0.554 | 0.935 | -0.167 | 0.374 | <.001 |
| GS | SRLQ 4 > Neutral | 238 | 0.449 | 0.561 | 0.947 | -0.127 | 0.505 | .004 |
| GS | SRLQ 5 > Neutral | 215 | 0.406 | 0.558 | 0.943 | -0.113 | 0.588 | .009 |
| ES | SRLQ 9 > Neutral | 232 | 0.438 | 0.558 | 0.941 | -0.135 | 0.492 | .002 |
| TMTS | SRLQ 10 > Neutral | 124 | 0.234 | 0.556 | 0.939 | -0.063 | 0.771 | .146 |

*Note*.

LC: learner-content engagement; SCI: social competencies with instructors; TC: technical competencies; ES: environment structuring; GS: goal setting; TMTS: time management and task strategies.

ESOQ questions: 22-‘Students use optional online resources to explore topics in more depth.’

OLRQ questions: 1-‘ I have a sense of self-confidence in using computer technologies for specific tasks’; 10-‘I am confident that I can timely inform the instructor when unexpected situations arise’; 11-‘I am confident that I can express my opinions to instructor respectfully’.

SRLQ questions: 3-‘I keep a high standard for my learning in my online courses’; 4-‘I set goals to help me manage studying time for my online courses’; 5-‘I don't compromise the quality of my work because it is online’; 9-‘I choose a time with few distractions for studying for my online courses’; 10-‘I try to take more thorough notes for my online courses’.

[a]: *p*-value of chi-square test.